\documentclass[seceq]{ptptex}

\usepackage{graphicx}

\notypesetlogo                       

\title{
Deformation around neutron-rich Cr isotopes in 
axially symmetric Skyrme-Hatree-Fock-Bogoliubov method
}

\author{Hiroshi \textsc{Oba}$^{1}$ , Masayuki \textsc{Matsuo}$^{2}$ }

\inst{
$^1$Graduate School of Science and Technology, Niigata University, Niigata
950-2181, Japan\\
$^2$Department of Physics, Faculty of Science, Niigata University, Niigata
950-2181, Japan
}

\abst{
We analyse deformation mechanism in neutron-rich Cr, Fe and Ti isotopes
 with $N$=32-44 
by means of a Skyrme-Hartree-Fock-Bogoliubov mean-field 
code employing a two-dimensional mesh representation 
in the cylindrical coordinate system.
Evaluating systematically the quadrupole deformation energy,
 we show that the Skyrme parameter set SkM* gives a quadrupole
 instability around the neutron number $N$$\sim$38-42 in Cr isotopes, where
 the deformation energy curve suggests a transitional behavior with a
 shallow minimum extending to a large prolate deformation.
Roles of a deformed $N$=38 gap and the position of the neutron
 $g_{\frac{9}{2}}$ orbit are analysed in detail. 
}

\begin{document}

\maketitle

\section{Introduction}
Deformation in the nuclear shape is one of the fundamental
features of the nuclear structure. Mechanism of the deformation 
reflects quantal effects such as the shell effect, the pairing
correlation, and the long range proton-neutron correlation.
Thus, analyses of a new deformation region often provide us with 
new insights into nuclear structure. A recent example is 
the quadrupole deformation in neutron-rich $Z$$\sim$12 isotopes, which 
have been a crucial probe to disclose the   
weakening of the $N$=20 shell gap in the neutron-rich
nuclei\cite{MG32}. More recently 
another possibility of a new deformation region has been suggested
in neutron-rich Cr isotopes\cite{Sorlin}\tocite{CR_RIKEN}. 

The self-consistent mean-field theory 
is a powerful scheme to analyse the deformation mechanism when
we deal with 
nuclei far from the stability line in medium and heavy mass
regions\cite{SUM_MEAN}. 
There are three important features which should be taken into account
in this case: the pairing correlation, the continuum states and the
deformation. 
The pairing correlation can be described by
using the Bogoliubov's generalized quasi-particle scheme while
the conventional BCS approximation is not suitable for nuclei
far from the stability line. The continuum states become important
in nuclei near the drip-lines
where the Fermi energy becomes close to zero. In this case one has to treat
continuum quasi-particle
states which have spatially extended wave functions. These two
requirements are fulfilled in the Hartree-Fock-Bogoliubov(HFB)
 formalism once it is represented in the coordinate-space\cite{1DHFB_DOBA,1DHFB_BUL}.
 To describe the deformation, then, the HFB method has to be extended to a
two- or three-dimensional problem, for which several techniques 
using the truncated Hartree-Fock
basis\cite{two_basis1,two_basis2,yamagami}, 
the transformed harmonic oscillator basis\cite{stoitsov,stoitsov2}, 
and the B-Spline basis\cite{oberacker,oberacker2} have been
developed recently. 

In the present paper, we develop our own 2D coordinate-space Skyrme-HFB 
code, which enables us to describe axially deformed unstable nuclei.
The formalism is similar to that of Ref.~\citen{oberacker}, which utilizes
 the B-Spline and Galerkin methods on top of the cylindrical coordinate
 system. In the present approach, we adopt a two-dimensional mesh representation 
using the cylindrical coordinate system and a multi-points formula for differential operators.
Since the 2D mesh approach using the
cylindrical coordinate has not been explored except a few
 Hartree-Fock\cite{negele} and Hartree-Fock-Bogoliubov
calculations\cite{yoshida,yoshida2}, we shall test numerical accuracy of our code.
Then, as an application of this code,
we shall investigate the possible deformation region around the neutron-rich
Cr isotopes. A global calculation of the ground state deformation
including the neutron-rich Cr isotopes is performed by means of the 
Skyrme-HFB\cite{stoitsov2,THO_SkMs,Goriely1} and the
relativistic mean-field models\cite{lalazissis}. 
Here we perform detailed analysis of the deformation mechanism
by looking into the deformation energy curve and the shell structure of
the single-particle orbits, and dependence on the Skyrme interactions.

\section{The HFB formalism}
\subsection{The HFB equation in the cylindrical coordinates system}
The HFB equation in the coordinate space representation is written
 as
\begin{equation}
 \left (
 \begin{array}{cc}
  h^{q}_{\sigma \sigma'}(\mathbf{r})-\lambda &
   \tilde{h}^{q}_{\sigma \sigma'}(\mathbf{r}) \\
  \tilde{h}^{q}_{\sigma \sigma'}(\mathbf{r}) &
   -(h^{q}_{\sigma \sigma'}(\mathbf{r})-\lambda)
 \end{array}
 \right )
 \left (
 \begin{array}{c}
  \phi_{\alpha q}^{(1)}(\mathbf{r}\sigma) \\
  \phi_{\alpha q}^{(2)}(\mathbf{r}\sigma)
 \end{array}
 \right )
 = E_{\alpha q}
 \left (
 \begin{array}{c}
  \phi_{\alpha q}^{(1)}(\mathbf{r}\sigma) \\
  \phi_{\alpha q}^{(2)}(\mathbf{r}\sigma)
 \end{array}
 \right )
\end{equation}
assuming the local HF potential $h^{q}_{\sigma \sigma'}(\mathbf{r})$ 
and the local pairing potential $\tilde{h}^{q}_{\sigma \sigma'}(\mathbf{r})$
for neutrons and protons ($q$ = n or p) with 
$\sigma$ and $\sigma'$ being spin indexes ($\sigma$,$\sigma$' = $\uparrow$,$\downarrow$).
In the present paper we assume the axial symmetry of nuclear
deformation and the $z$-axis is chosen as the symmetry axis.
The quasi-particle wave function has a quantum number $\Omega$ for the
$z$-component of the total angular momentum.
Using the cylindrical coordinate $(r,z,\varphi)$,
the quasi-particle wave function of the $n$-th eigenstate is written as 
\begin{eqnarray}
\left (
\begin{array}{c}
\phi_{n \Omega q}^{(1)}(\mathbf{r}\sigma) \\
\phi_{n \Omega q}^{(2)}(\mathbf{r}\sigma)
\end{array}
\right )
 &=&
\frac{1}{\sqrt{2\pi}}
\left (
 \begin{array}{c}
  e^{i(\Omega-\frac{1}{2}) \varphi} \phi_{n \Omega q}^{(1)\uparrow}(r,z) \\
  e^{i(\Omega+\frac{1}{2}) \varphi} \phi_{n \Omega q}^{(1)\downarrow}(r,z) \\
  e^{i(\Omega-\frac{1}{2}) \varphi} \phi_{n \Omega q}^{(2)\uparrow}(r,z) \\
  e^{i(\Omega+\frac{1}{2}) \varphi} \phi_{n \Omega q}^{(2)\downarrow}(r,z)
 \end{array}
\right ).
\end{eqnarray} 
Expressing the Hearted-Fock Hamiltonian $h(r,z,\varphi)$ 
and the pairing Hamiltonian $\tilde{h}(r,z,\varphi)$ as
\begin{equation}
 h^{q}(r,z,\varphi) =
 \left (
 \begin{array}{cc}
  h^{q}_{\uparrow \uparrow}(r,z) & e^{-i\varphi}h^{q}_{\uparrow \downarrow}(r,z) \\
  e^{i\varphi}h^{q}_{\downarrow \uparrow}(r,z) & h^{q}_{\downarrow \downarrow}(r,z)
 \end{array}
 \right ),
\end{equation}
\begin{equation}
 \tilde{h}^{q}(r,z,\varphi) =
 \left (
 \begin{array}{cc}
  \tilde{h}^{q}_{\uparrow \uparrow}(r,z) & e^{-i\varphi}\tilde{h}^{q}_{\uparrow \downarrow}(r,z) \\
  e^{i\varphi}\tilde{h}^{q}_{\downarrow \uparrow}(r,z) & \tilde{h}^{q}_{\downarrow \downarrow}(r,z)
 \end{array}
 \right ),
\end{equation}
the HFB equation in the cylindrical coordinate representation is written as
\begin{equation}
  H_{q}\Phi_{n \Omega q} = E_{n \Omega q} \Phi_{n \Omega q} \label{cyhfbeq}
\end{equation}
 where 
\begin{equation}
 H_{q} = \left (
 \begin{array}{cccc}
  h^{q}_{\uparrow \uparrow}(r,z)-\lambda & h^{q}_{\uparrow \downarrow}(r,z) &
  \tilde{h}^{q}_{\uparrow \uparrow}(r,z) & \tilde{h}^{q}_{\uparrow \downarrow}(r,z) \\
  h^{q}_{\downarrow \uparrow}(r,z) & h^{q}_{\downarrow \downarrow}(r,z)-\lambda &
  \tilde{h}^{q}_{\downarrow \uparrow}(r,z) & \tilde{h}^{q}_{\downarrow \downarrow}(r,z) \\
  \tilde{h}^{q}_{\uparrow \uparrow}(r,z) & \tilde{h}^{q}_{\uparrow \downarrow}(r,z) &
  -h^{q}_{\uparrow \uparrow}(r,z)+\lambda & -h^{q}_{\uparrow \downarrow}(r,z) \\
  \tilde{h}^{q}_{\downarrow \uparrow}(r,z) & \tilde{h}^{q}_{\downarrow \downarrow}(r,z) &
  -h^{q}_{\downarrow \uparrow}(r,z) & -h^{q}_{\downarrow \downarrow}(r,z)+\lambda 
 \end{array}
 \right )
\end{equation}
 and 
\begin{equation}
 \Phi_{n \Omega q} = \left (
  \begin{array}{c}
   \phi_{n \Omega q}^{(1)\uparrow}(r,z) \\
   \phi_{n \Omega q}^{(1)\downarrow}(r,z) \\
   \phi_{n \Omega q}^{(2)\uparrow}(r,z) \\
   \phi_{n \Omega q}^{(2)\downarrow}(r,z)
  \end{array}
 \right ) .
\end{equation}
We adopt the Skyrme force and  $h^{q}_{\sigma\sigma'}(r,z)$ is
expressed\cite{oberacker} in terms of the 
force parameters and the normal density $\rho_{n,p}(r,z)$, the kinetic density
$\tau_{n,p}(r,z)$, the spin-orbit density $J_{n,p}(r,z)$.
We adopt the density dependent delta interaction(DDDI)
\[
 v_{pair}=\frac{1}{2}V_{0}(1-P_{\sigma})
 \left( 1- \left ( \frac{\rho (\mathbf{r})}{\rho _{0}} \right)^{\eta} \right)
 \delta(\mathbf{r}-\mathbf{r'}) \label{pair_force}
\]
for the effective interaction producing the pairing potential
$\tilde{h}^{q}_{\sigma\sigma'}$, where $\tilde{h}^{q}_{\sigma\sigma'}$
is expressed in terms of the total nucleon density $\rho(r,z)$ and the pair density $\tilde{\rho}_{q}(r,z)$.
The eigenstates and eigenenergies of the quasi-particles are obtained by
solving the HFB equation (\ref{cyhfbeq}) for each $\Omega$.
Since we assume the time reversal symmetry,
we need to solve only for $\Omega>0$.

We consider a rectangular area $0 \le r \le r_{max}$ and $-z_{max} \le z
\le z_{max}$ in the $(r,z)$ plane.
The $r$-axis is discretized at $r=h/2, 3h/2, \cdots, r_{max}$
with $N_{r}$ mesh points,
while the $z$-axis is discretized from $-z_{max}=-N_{z}h$ 
to $z_{zmax}=N_{z}h$ with the same equidistant interval $h$.
The quasi-particle wave functions $\phi(r,z)$(the indices are omitted
here for simplicity) and the HFB Hamiltonian $H_{q}$ are 
represented on this two-dimensional mesh.
At the outer boundary $z= \pm z_{max}$ and $r=r_{max}$, we impose $\phi(r_{max},z)=0$
and $\phi(r,\pm z_{max})=0$.
At the origin $r=0$, we impose $ \phi(-r,z) = (-)^{l_{z}}\phi(r,z)$
 determined by the parity $(-)^{l_{z}}$
 where $l_{z}$ is the $z$-component of the orbital angular momentum. 
Note that we utilize shifted mesh points $r=(n+\frac{1}{2})h$
 so that we do not need to evaluate the value of $\phi$ at $r=0$.
We made a comparison with calculations where 
the $r$-axis is discretized on a slightly different mesh points  
$r=0, h, 2h, \cdots, r_{max}$ and we found that 
the adopted discretization with  $r=(n+\frac{1}{2})h$ has better accuracy.
We adopt the 11-points formula to represent differential operators.
Concerning the integrals, we adopt trapezoidal formula since
its is known that the high order formula does not always guarantee a better 
accuracy\cite{9point}.
Consequently a quasi-particle wave function is 
expressed as a vector with $N=4N_{z}N_{r}$ dimension 
and the HFB Hamiltonian is a matrix having the size of $N^{2}$.
We obtain the eigenvalues and the eigenfunctions by diagonalizing this 
Hamiltonian matrix.
The Hamiltonian is non-symmetric because of the finite-points
representation of the differential operators around $r=0$.
We diagonalize the Hamiltonian using the QR method  
to solve the non-symmetric eigenvalue problem.

Since we use the delta type pairing force,
we introduce a cut-off with respect to the quasi-particle energy.
The adopted value of cut-off energy is 60 MeV.
We need also a cut-off with respect to the azimuthal quantum number $\Omega$,
and we adopted the cut-off value of $\Omega = \frac{15}{2}$.
In order to obtain the self-consistent solution, we utilize an iteration
scheme. As an initial condition, we start with  
single-particle orbits in a deformed Woods-Saxon potential. 
As an initial pairing potential, we chose one having a Woods-Saxon shape.
We revise only a fraction of the densities in each step of iteration, 
and this fraction is randomized within an interval 0.2-0.4.
When about 200 iterations pass, the increment of the energy reaches
$1\times10^{-5}$ MeV, at which we stop the iteration.
In order to obtain a deformation energy, we perform also a constrained HFB
calculation.
As a constraint operator, we use the mass quadrupole 
operator $\hat{Q}_{2}$, which is defined as 
\[
\hat{Q}_{2}=\sqrt{\frac{16\pi}{5}} R^{2} Y_{20} (\hat{r})
= 2z^{2}-r^{2}
\]
($R=\sqrt{r^2+z^2}$). We use the quadratic constraint method\cite{RingSchuck}. 

\subsection{Test of the numerical code}

\begin{figure}[h]
  \begin{center}
    \begin{tabular}{cc}
     \includegraphics[width=4cm]{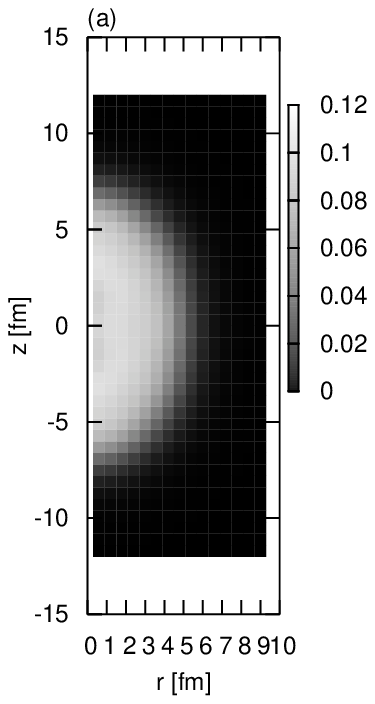} &
     \includegraphics[width=4cm]{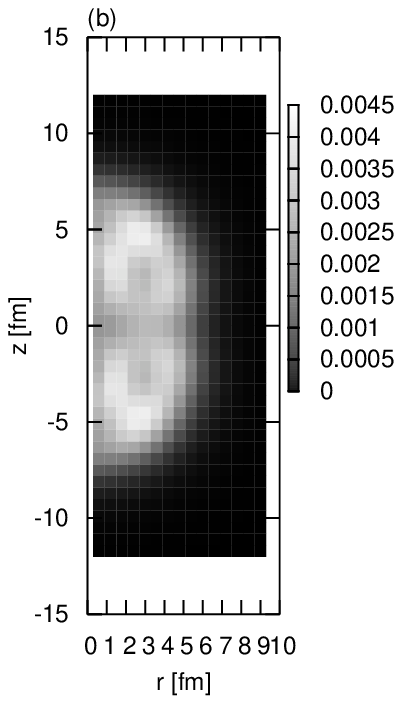} 
    \end{tabular}
   \caption{(a) The density $\rho_{n}(r,z)$
    and (b) the pair density $\tilde{\rho}_{n}(r,z)$ for neutrons in $^{102}$Zr. 
   The unit in the legend is fm$^{-3}$.
   } \label{Zr102neutron2D}
  \end{center}
\end{figure}

To check the accuracy of the code, we compare our result with 
those in the two previous Skyrme-HFB calculations.
One is the THO basis method developed by Stoitsov et
al.\cite{stoitsov} and the other is
 the B-Spline basis method developed by Ter\'{a}n et al.\cite{oberacker}.     
We choose neutron-rich Zr isotopes.
Along with Refs.\citen{stoitsov} and \citen{oberacker},
we adopt the Skyrme parameter SLy4 and the volume type pairing force,
$i.e.$ $\eta=0$.
The box size is 12 fm for both $r_{max}$ and $z_{max}$, and
the mesh interval $h$ is 0.6 fm. 
The pairing strength $V_{0}$ is the same in Refs.~\citen{oberacker} and \citen{oberacker2}.
The obtained ground state properties of neutron-rich Zr 
isotopes are shown in 
Table \ref{Zr102_table} and Fig.\ref{Zr102neutron2D}. 
\begin{table}[h]
\caption{The total binding energy BE, the Fermi energy $\lambda_{q}$,
the average pairing gap $\Delta_{q}$ and the RMS radius $\sqrt{\langle r^{2}
 \rangle}$ in $^{102}$Zr obtained in the present work.
 The values\cite{oberacker} of the B-Spline and the
 THO methods are compared.
 }\label{Zr102_table}
\begin{center}
\begin{tabular}{|c|c|c|c|}\hline
& present work & B-Spline & THO \\ \hline
BE [MeV] &  -856.92 & -859.61 & -859.40 \\ \hline
$\lambda_{n}$ [MeV] & -5.42 & -5.46 & -5.42 \\ \hline
$\lambda_{p}$ [MeV] & -12.07 & -12.08 & -12.10 \\ \hline
$\Delta_{n}$ [MeV] & 0.23 & 0.31 & 0.56 \\ \hline
$\Delta_{p}$ [MeV] & 0.35 & 0.34 & 0.62 \\ \hline
$\sqrt{\langle r^{2} \rangle}$ [fm] & 4.58 & 4.58 & 4.58 \\ \hline
\end{tabular}
\end{center}
\end{table}
The density and the pair density for neutrons are shown in 
Fig.\ref{Zr102neutron2D}.
It is seen that the neutron density has a large quadrupole
deformation and the neutron pair density is also deformed.
Note that the pair density exhibits a complex structure in the
 inner region.

In Table \ref{Zr102_table}, we compare our result for $^{102}$Zr with
the results of the B-Spline and the THO methods in Ref.~\citen{oberacker}.
It is seen that our code reproduces reasonably well. 
Especially, the root mean square (RMS) radius $\sqrt{\langle r^{2}
\rangle}$ and the Fermi energies $\lambda_{n,p}$ are in a good agreement. 
Note however that there is a deficiency in the total binding energy 
by about 2-3 MeV.
This is probably due to use of the moderate mesh size ($h$=0.6 fm) and 
the finite-points formula for the differential operators 
as a deficiency due to the finite mesh size has been
known in previous HF+BCS calculations\cite{Bonche,Baye}.
There is some deviation in the average neutron pairing gap $\Delta_{n}$.
It should be noted that the pairing gap is rather small $\sim$ 300 keV.
Generally the value of the pairing gap is
sensitive to minor differences, especially when the pairing gap is small.
A small difference in defining the cut-off
energy may be suspected 
as we adopt the quasi-particle energy cut-off while the THO and B-Spline methods
use so called the equivalent single particle energy cut-off.
The finite mesh size effect could be another origin.

\begin{figure}[h]
  \begin{center}
    \begin{tabular}{cc}
     \includegraphics[width=6cm]{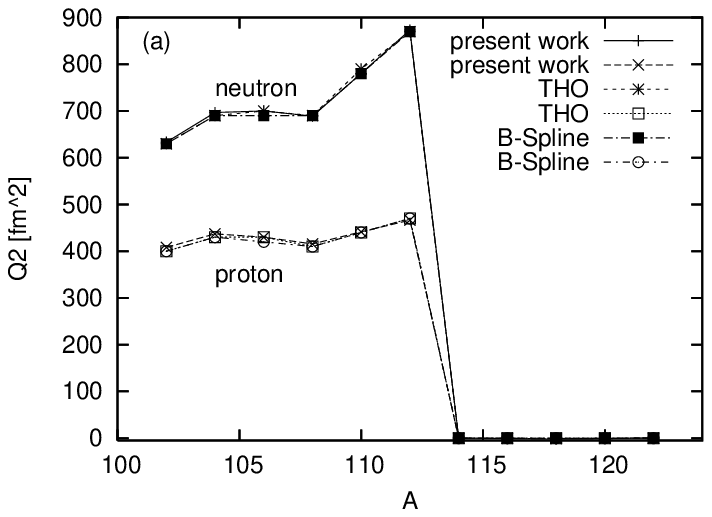} &
     \includegraphics[width=6cm]{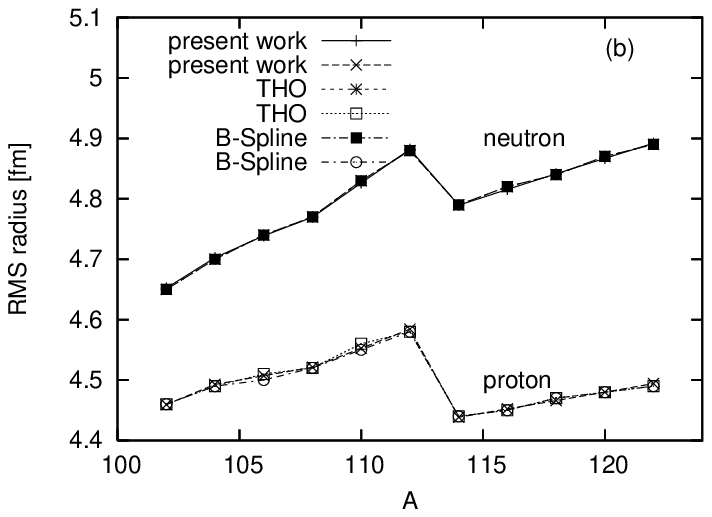} \\
     \includegraphics[width=6cm]{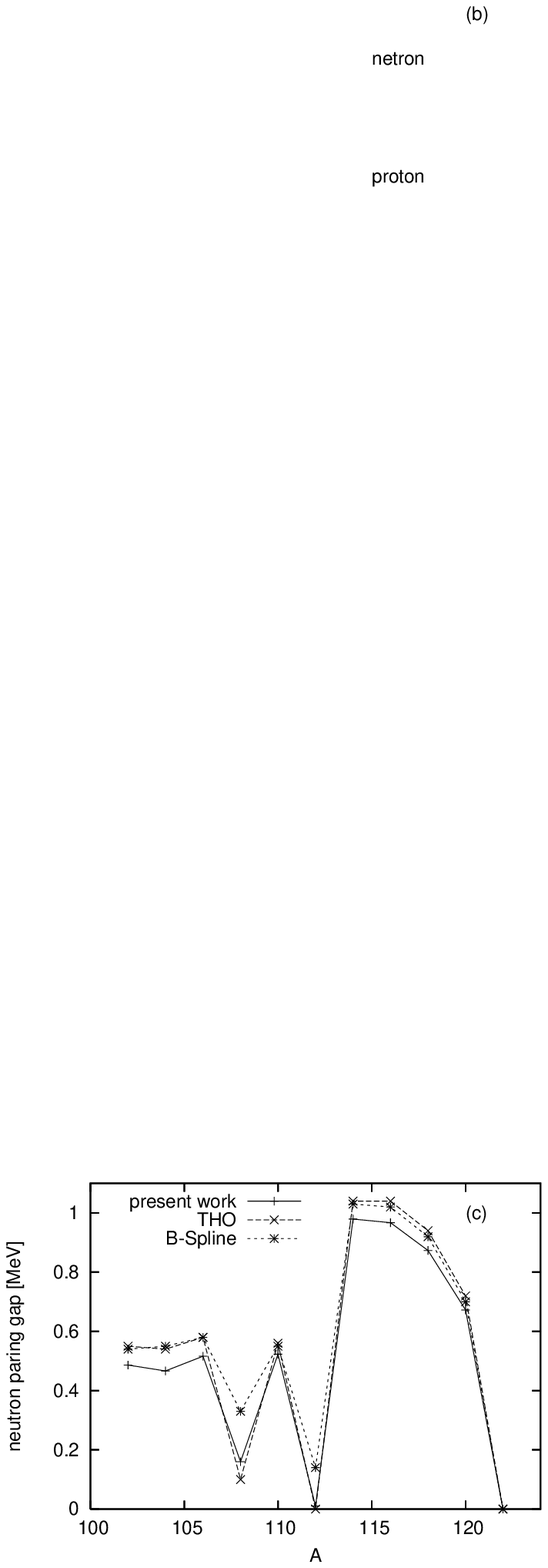} &
     \includegraphics[width=6cm]{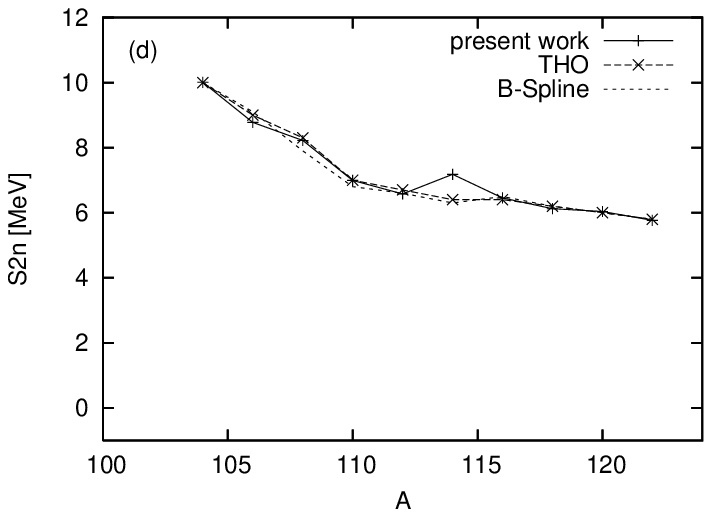}
    \end{tabular}
     \caption{The comparison of (a) the mass quadrupole moment $Q_{2}$, 
(b) the RMS radius $\sqrt{\langle r^{2}
   \rangle}$, (c) the neutron average pairing gap $\Delta_{n}$, and (d) the two
   neutron separation energy $S_{2n}$ in the neutron-rich Zr isotopes
   between the present work and the B-Spline and the THO
   methods\cite{oberacker2}.}\label{Zr_chain}
  \end{center}
\end{figure}

To make a systematic comparison,
we have calculated an isotopic chain of zirconium for $N$=102-122.
The mass quadrupole moment
$Q_{2}$, the RMS radius, the neutron and proton pairing gaps, and the two-neutron separation
energy are compared with those of 
Ref.~\citen{oberacker2} in Fig.\ref{Zr_chain}(a)-(d).
In this calculation, the pairing strength is 
$V_{0}$=-187.1305 MeV fm$^{-3}$.
The quadrupole moment (Fig.\ref{Zr_chain}(a)) as well as the RMS radius (Fig.\ref{Zr_chain}(b)) 
agree very well with those in the B-Spline and the THO methods in the whole
region ($A$=102-122) shown in the figure. 
Especially the deformation is described  well.
The difference in the neutron separation energy is smaller than the 2-3 MeV difference in the absolute value
of the total binding energy.
The better accuracy in the relative quantities is consistent with the
same feature found in the previous Skyrme-HF + BCS calculations using 
the 3D Cartesian mesh representation \cite{tajima}. 
The average pairing gap is also reproduced reasonably well as shown in
Fig.\ref{Zr_chain}(c) although 
a small but non-negligible difference is noted.

\begin{figure}[h]
  \begin{center}
     \includegraphics[width=8cm]{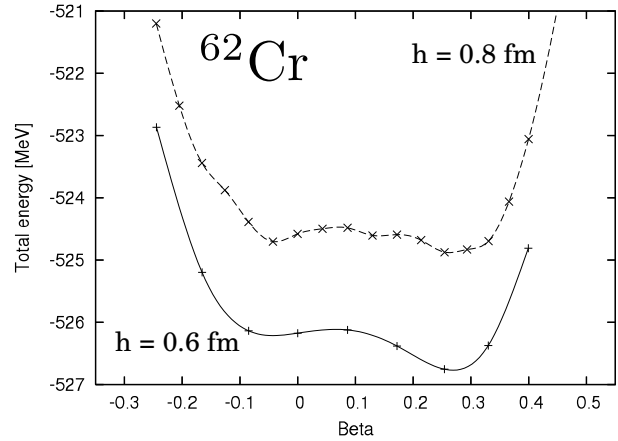}
   \caption{The deformation energy curves as a function of the
   deformation parameter $\beta$ calculated for $^{62}$Cr using SkM*.  
   The solid line is the case of $h=$0.6 fm and the dash line for 0.8 fm.\label{com06-08_Q2}}
  \end{center}
\end{figure}

In Figure \ref{com06-08_Q2}, we display the deformation energy curve as a
function of the quadrupole deformation parameter $\beta$ defined by 
$\beta = \sqrt{\frac{\pi}{5}}\frac{\langle \hat{Q_{2}} \rangle}{\langle r^{2} \rangle A }$. 
The dependence of the quadrupole deformation 
energy on the mesh size is seen from the comparison of the results obtained with 
the mesh sizes 0.6 and 0.8 fm.
If we shift the total energy by about 1.8 MeV, we see that the difference
between the two curves becomes within about 200 keV.
We thus conclude that the deformation energy can be evaluated to the
accuracy of this order.

\section{Quadrupole deformation around the neutron-rich Cr isotopes with
 $N\sim$38}
\subsection{The neutron-rich Cr isotopes and neutron single-particle gaps}
The observed $2^{+}_{1}$ energy of the neutron-rich Cr isotopes 
decreases monotonically with increasing the neutron number 
from $E_{2}$ = 1007 keV in $^{56}$Cr down to 446 keV in
$^{62}$Cr\cite{Sorlin}, 
suggesting a new region of deformation.
The energy ratio $E_{4}/E_{2}$ between the $2_{1}^{+}$ and $4_{1}^{+}$ 
excited states becomes large $E_{4}/E_{2}$=2.65 in
$^{62}$Cr\cite{CR_RIKEN} while the ratio is small $E_{4}/E_{2} \sim
2.2$ for lighter isotopes\cite{zhu}. 
The proton inelastic scattering experiment\cite{CR_RIKEN} indicates that the
$2_{1}^{+}$ state accompanies a large quadrupole deformation $\beta\sim 0.25$. 
Shell model analyses\cite{Sorlin,caurier} suggest 
a possible deformation for $N\ge$38 as an extended model space
including at least the neutron $1g_{\frac{9}{2}}$ and $2d_{\frac{5}{2}}$ orbits in the higher shell 
is necessary to describe the decreasing trend of $E_{2}$ with increasing
the neutron number. 
There are several self-consistent mean-field calculations
\cite{THO_SkMs,Goriely1,lalazissis} for the systematics of
the ground state deformation, including the neutron-rich Cr isotopes. 
The calculation using the Skyrme parameter set SkM*\cite{THO_SkMs} produces a large 
prolate deformation around $^{62}$Cr, but the other Skyrme models using
SLy4\cite{THO_SkMs}, SkP\cite{THO_SkMs}, BSk2\cite{Goriely1} and 
the relativistic mean-field model using NL-SH\cite{lalazissis} do not. 
Here we analyse the microscopic deformation mechanism with focuses on
the parameter set SkM*. 

First, we discuss the deformation energy curve in $^{56-68}$Cr.
\begin{figure}[h]
  \begin{center}
     \includegraphics[width=8cm]{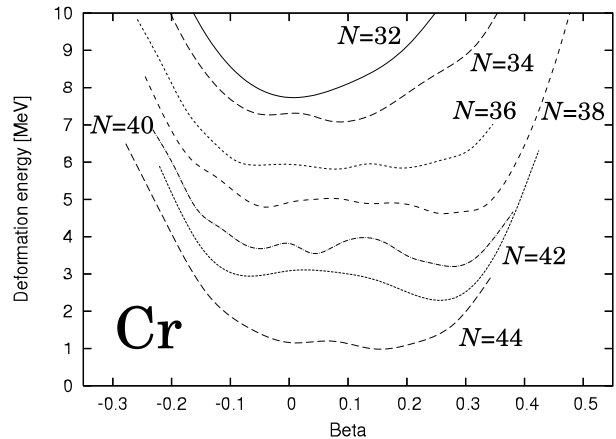} 
   \caption{The quadrupole deformation energy curves of the neutron-rich Cr
   isotopes calculated with SkM*. 
   The energy is arbitrarily shifted for the sake of comparison.}\label{Cr_isotope}
  \end{center}
\end{figure}
We adopt SkM* and the volume type pairing force ($\eta = 0$ in (\ref{pair_force}))
with the pairing strength $V_{0}$ = -200 MeV fm$^{-3}$ .
The value of $V_{0}$ is adjusted to reproduce the odd-even mass
difference in $^{56-66}$Cr, which have the values between 1.5 and 2.0 MeV. 
The ground state values of $\beta$, $\Delta_{n}$ and $\Delta_{p}$ are shown
in Table \ref{gf_occ}.
The calculated deformation energy curves of the neutron-rich Cr isotopes
are plotted in Fig.\ref{Cr_isotope}.
It is seen that the deformation energy curve becomes soft toward 
the prolate direction with increasing the neutron number from $N$=32.
In $^{60}$Cr the deformation energy is almost flat between $\beta$=-0.1 and 0.3.
In $^{62,64,66}$Cr($N$=38-42), the minimum with a large quadrupole deformation appears
around $\beta$=0.25 - 0.3, but then
the deformation of the minimum decreases for $N\ge$42.
The calculated isotopic tendency is consistent with the observed trend
of $E_{2}$ from $N$=32 to $N$=38.
It is noted, however, that the energy difference between
the spherical configuration($\beta$=0) and the quadrupole deformed
minimum is small. In the case of $^{62}$Cr, the difference is about
290 keV. If we estimate the zero-point energy in terms of the experimental value
of $E_{2}$=446 keV, the deformation energy and the zero-point
energy have comparable magnitudes. 
Thus it is suggested, within the SkM* model, that even in $^{62-64}$Cr exhibiting the
largely deformed minima a well developed static
deformation is not realized, and that a large amplitude quadrupole motion
of a transitional character is expected.
The transitional nature seems to be consistent with the observed
$E_{4}/E_{2}$ ratio\cite{CR_RIKEN} 2.65 in $^{62}$Cr which lies in between 
the vibrator and the rotor limits 2 and 3.33.

\begin{figure}[h]
  \begin{center}
    \begin{tabular}{cc}
     \includegraphics[width=6cm]{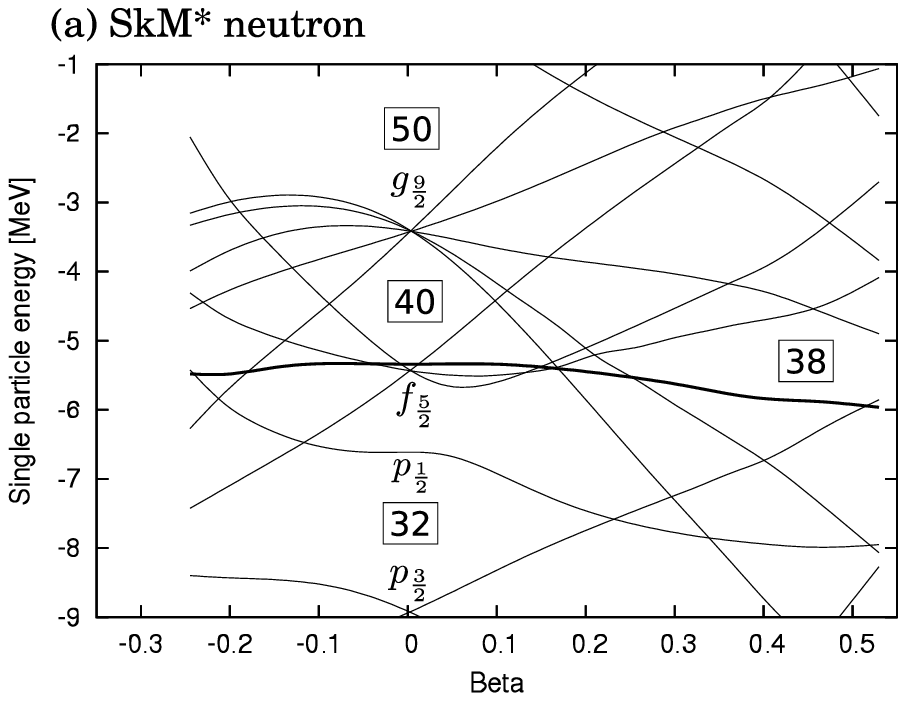} &
     \includegraphics[width=6cm]{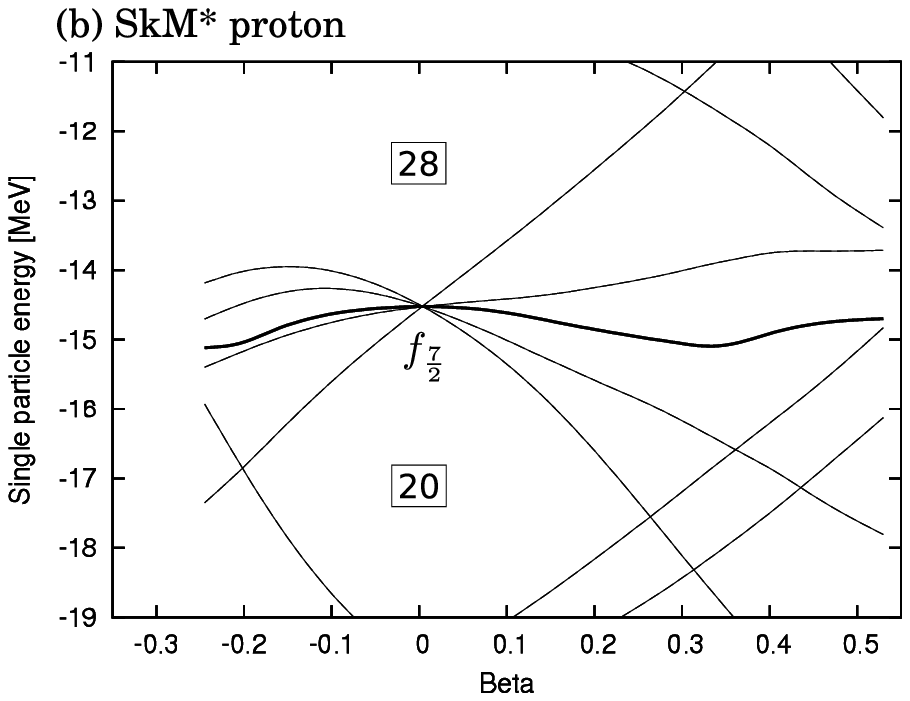} \\
     \includegraphics[width=6cm]{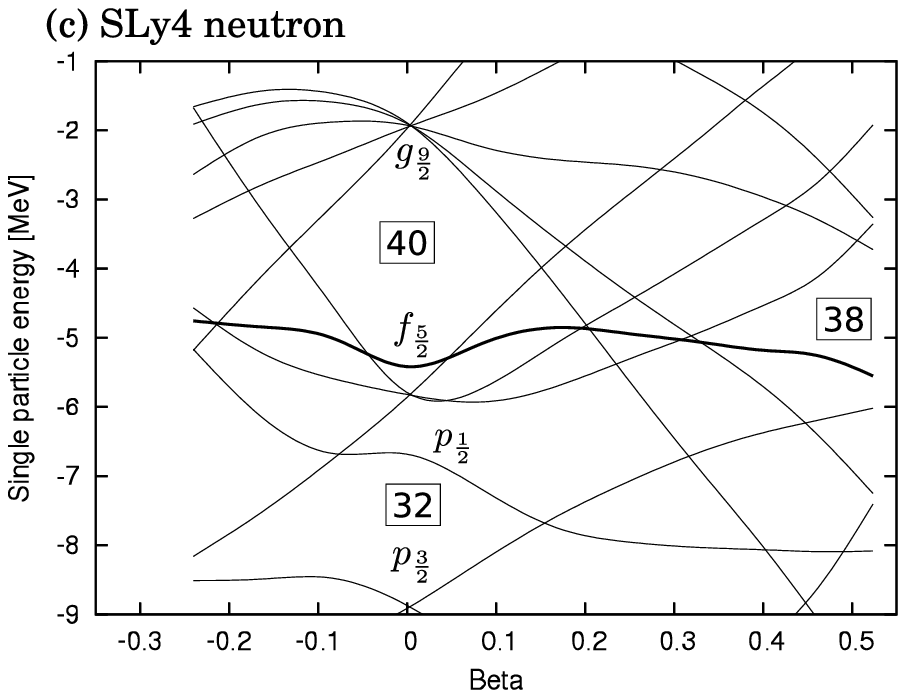} &
     \includegraphics[width=6cm]{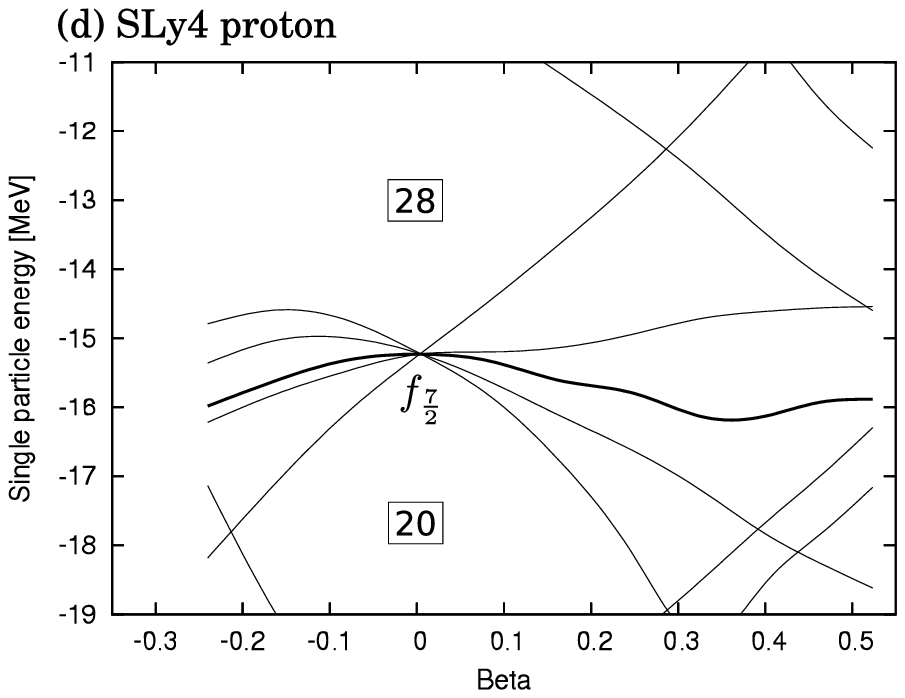}
    \end{tabular}
   \caption{The single-particle energies obtained with SkM*(left panels (a)(b)) and
   SLy4(right panels (c)(d)) for $^{62}$Cr. The top panels (a)(c) are for
   neutrons, and the bottom panels (b)(d) are for 
   protons. The thick solid line is the Fermi energy.  
   The horizontal axis is the quadrupole deformation parameter
   $\beta$.}\label{Cr62_sp_energy}
  \end{center}
\end{figure}

Next we analyse the mechanism of the deformation.
For this purpose, we plot the Hartree-Fock single-particle energies 
as a function of $\beta$.
The single-particle orbits are obtained by rediagonalizing the HF
Hamiltonian associated with the constrained Skyrme-HFB solution.
The result is shown in Fig.\ref{Cr62_sp_energy}(a) and (b).
The neutron subshell gap $N$=40 between the 
$f_{\frac{5}{2}}$ and $g_{\frac{9}{2}}$ orbits in the spherical region 
and an energy gap $N$=38 in the prolate region around $\beta$=0.25-0.5 
are notable.
As the magnitude of the deformed $N$=38 gap is comparable to the spherical $N$=40 subshell
 gap, 
it is easy to deduce that the deformed $N$=38 gap may drive the nucleus toward
 prolate deformation with $\beta \sim$0.3. 
We also note that the neutron $g_{\frac{9}{2}}$ orbit plays an important role 
in forming the deformed states.
Especially the $\Omega = \frac{1}{2}$ and $\frac{3}{2}$
orbits which stem from the deformation splitting of $\nu
g_{\frac{9}{2}}$ are relevant as they exhibit the steepest down-sloping for $\beta \ge 0$.
If neutrons occupy these $\nu g_{\frac{9}{2}}$ orbits, they cause 
a strong driving force toward prolate deformation. 
This situation can be realized for $\beta \gtrsim 0.20$. 
We show in Table \ref{gf_occ} the occupation number for all 
of the $\nu g_{\frac{9}{2}}$ orbits.
The occupation number $\sim$3.0-3.8 in $^{62}$Cr and $^{64}$Cr, 
where the largest ground state deformation is realized, 
is consistent with the interpretation that the $\nu g_{\frac{9}{2}}$ 
 $\Omega = \frac{1}{2}$ and $\frac{3}{2}$ orbitals are largely occupied. 
As the neutron number goes beyond $N$=38, the occupation number of 
$\nu g_{\frac{9}{2}}$ still increases slightly (3.8 in $^{64}$Cr with
$N$=40 and 4.4 in $^{66}$Cr with $N$=42). 
However, the quadrupole deformation then turns to decrease at $N$=42 since
the upsloping orbits [303 $\frac{5}{2}$] and
[301$\frac{1}{2}$] stemming from $\nu f_{\frac{5}{2}}$ and 
$\nu p_{\frac{1}{2}}$ are occupied. 
This explains the reason for the largest quadrupole deformation in
$^{62,64}$Cr with $N$=38-40. 
The important role of the occupation of $\nu g_{\frac{9}{2}}$ is
consistent with the conclusions of the shell model 
analyses\cite{Sorlin,freeman,zhu,caurier}.

\subsection{Neutron-rich Ti and Fe isotopes}
\begin{figure}[h]
  \begin{center}
    \begin{tabular}{cc}
     \includegraphics[width=6cm]{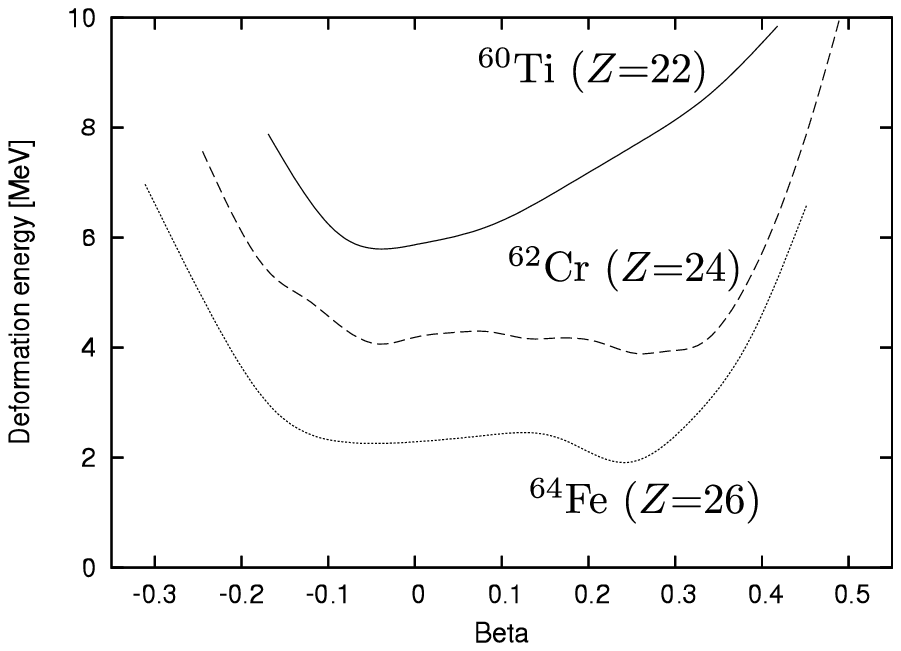} &
     \includegraphics[width=6cm]{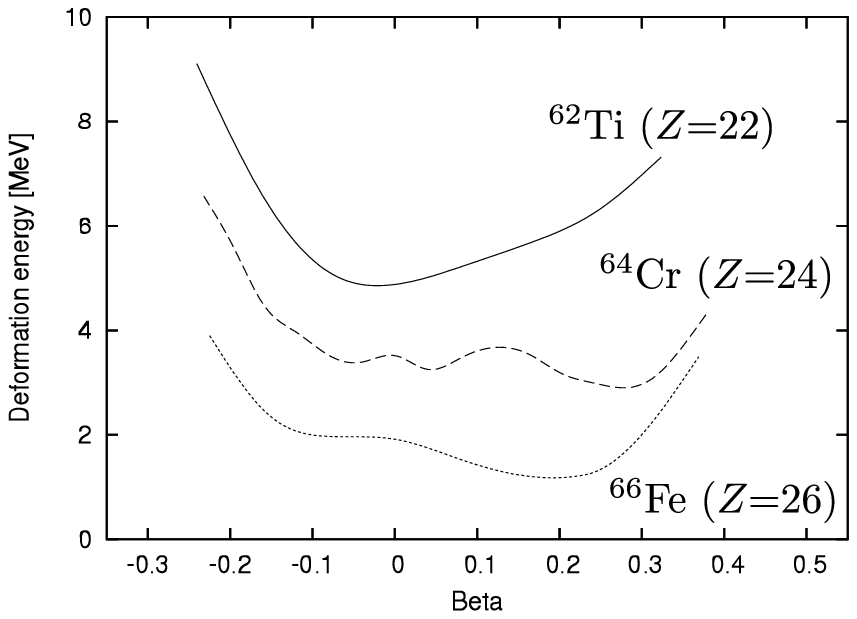} 
    \end{tabular}
   \caption{The quadrupole deformation energy curves of the neutron-rich
   $N$=38 and 40 isotones around $Z$=24 obtained with SkM*.}\label{Cr_Fe_Ti}
  \end{center}
\end{figure}

It is interesting to investigate neutron-rich Fe and Ti isotopes
around $^{62,64}$Cr with $N$=38, 40 with which the largest deformation
is realized in the Cr isotopes.
Figure \ref{Cr_Fe_Ti} shows the deformation energy curve
 calculated with SkM* for the $N$=38 isotones 
$^{60}$Ti, $^{62}$Cr, $^{64}$Fe and for the $N$=40 isotones $^{62}$Ti,
$^{64}$Cr, $^{66}$Fe.
It is seen that the Cr and Fe isotopes have deformed minima at prolate
deformation while the Ti isotopes do not exhibit this feature. 
Since the importance of proton configurations is easily inferred, 
we look into the proton Nilsson diagram in Fig.\ref{Cr62_sp_energy}(b), which is for
$^{62}$Cr but essentially the same for the nuclei under discussion.
We also show the occupation number in the proton $f_{\frac{7}{2}}$
orbits in Table \ref{gf_occ}.
\begin{table}[h]
 \caption{The occupation numbers of the neutron $g_{\frac{9}{2}}$ orbit
 and the proton $f_{\frac{7}{2}}$ orbit, the average neutron and proton
 pairing gaps $\Delta_{n}$, $\Delta_{p}$ and the quadrupole
 deformation parameter $\beta$ associated with the HFB solutions obtained with SkM*.
}\label{gf_occ}
 \begin{center}
  \begin{tabular}{|c|c|c|c|c|c|}\hline
   & neutron $g_{\frac{9}{2}}$ & proton $f_{\frac{7}{2}}$ & $\Delta_{n}$ [MeV]
   & $\Delta_{p}$ [MeV] & $\beta$ \\ \hline
   $^{60}$Ti & 1.28  & 2.02  & 2.042 & 1.198 & 0.0   \\ \hline
   $^{62}$Ti & 1.28  & 2.02  & 2.096 & 1.157 & 0.0   \\ \hline
   $^{58}$Cr & 0.355 & 3.98  & 1.619 & 1.427 & 0.0   \\ \hline
   $^{60}$Cr & 0.794 & 3.98  & 1.970 & 1.388 & 0.0   \\ \hline
   $^{62}$Cr & 2.98  & 4.00  & 1.908 & 0.405 & 0.257 \\ \hline
   $^{64}$Cr & 3.85  & 4.02  & 1.904 & 0.212 & 0.262 \\ \hline
   $^{66}$Cr & 4.39  & 4.00  & 1.833 & 0.280 & 0.247 \\ \hline
   $^{68}$Cr & 5.40  & 4.00  & 1.756 & 0.456 & 0.222 \\ \hline
   $^{64}$Fe & 2.58  & 5.99  & 1.999 & 0.280 & 0.212 \\ \hline
   $^{66}$Fe & 3.76  & 6.00  & 1.951 & 0.008 & 0.231 \\ \hline
  \end{tabular}
 \end{center}
\end{table}
The occupation numbers are
$\sim$2, 4, and 6 in Ti, Cr and Fe respectively as naturally
expected from Fig.\ref{Cr62_sp_energy}(b).
Since the $\Omega=\frac{1}{2}$ and $\Omega=\frac{3}{2}$ orbits stemming
from $\pi f_{\frac{7}{2}}$ have steep slopes in the prolate direction,
 simultaneous occupation of these orbits in Cr gives a
 large driving force toward the prolate deformation.
We can expect that the deformation is a less favored in Fe than
in Cr because 
protons in Fe occupy the $\Omega = \frac{5}{2}$ orbits which is slightly
up-sloping.
Indeed the value of $\beta$ at the minima in $^{64,66}$Fe is little
smaller than those in $^{62,64}$Cr(cf. Table \ref{gf_occ} , Fig.\ref{Cr_Fe_Ti}).
On the other hand the spherical minima in Ti isotopes indicate that 
occupation of the $\Omega = \frac{1}{2}$ orbit alone is not enough to
cause deformation.

The observed $2_{1}^{+}$ energies\cite{Hannawald} in $^{64,66}$Fe are 
lower than that in $^{62}$Fe, and the observed ratio\cite{hoteling} $E_{4}/E_{2}$=2.36 
in $^{64}$Fe is smaller than that(=2.65) in $^{62}$Cr\cite{CR_RIKEN}. 
The deformation energy curves indicating smaller quadrupole deformation
in $^{64,66}$Fe than in $^{62,64}$Cr are consistent with these
experimental trends. In addition, the calculation predicts much less collectivity in
Ti than in Cr and Fe isotopes.
This is also consistent with higher $E_{2}$ in $^{58}$Ti than in $^{60}$Cr\cite{CR_RIKEN}.

\subsection{Relation to the $N \sim Z$=38-40 deformed region}
\begin{figure}[h]
  \begin{center}
     \includegraphics[width=8cm]{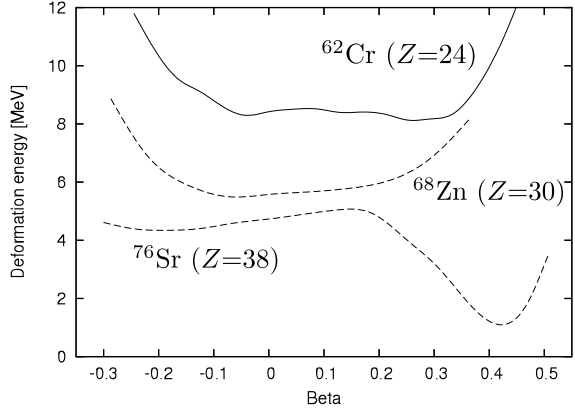}  
   \caption{The quadrupole deformation energy curves in the N=38 isotones obtained with SkM*.}\label{Cr_Zn_Sr_energy}
  \end{center}
\end{figure}

The role of the $N$=38 deformed shell gap reminds us of the mechanism of
the prolate deformation in proton-rich Sr and Zr isotopes with
$N \sim Z$=38-40 as the large deformation in the proton-rich $N$=$Z$ region
originates from the presence of a deformed shell gap at $N,Z$=38
 and 40 around $\beta \sim$ 0.4\cite{Bonche,NAZ_NeZ,HEY_NeZ}. 
It is therefore interesting to investigate how the deformation of the
neutron-rich Cr, Fe nuclei
is related to the $N \sim Z$=38-40 cases. 
Figure \ref{Cr_Zn_Sr_energy} compares the deformation energy curve 
in $^{62}$Cr with those of $^{68}$Zn($Z$=30) and $^{76}$Sr($Z$=38) 
in the $N$=38 isotone chain up to $N$=$Z$=38.
Note that the deformation energy curve in the $N$=$Z$=38 nucleus 
$^{76}$Sr has a well-developed deformed minimum around $\beta \sim$0.4.

\begin{figure}[h]
  \begin{center}
     \includegraphics[width=8cm]{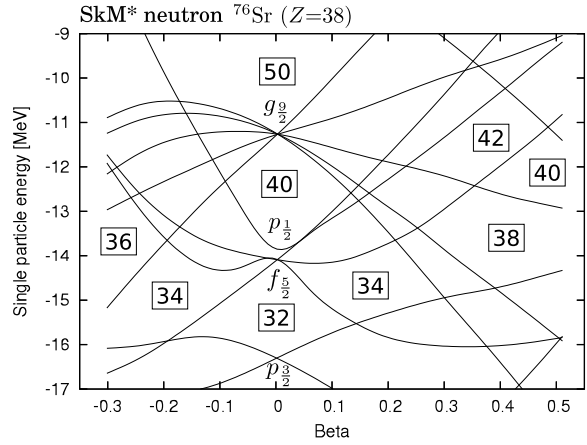}
   \caption{The neutron Nilsson diagram of $^{76}$Sr obatained with SkM*.}\label{Sr76_n_sp_energy}
  \end{center}
\end{figure}
Figure \ref{Sr76_n_sp_energy} plots the neutron Nilsson diagram in
$^{76}$Sr.
(The proton Nilsson diagram is not shown here as it is very similar 
to the neutron's except an overall energy shift due to the Coulomb
potential.)
Comparing the neutron Nilsson diagrams in Fig.\ref{Sr76_n_sp_energy}
 and in Fig.\ref{Cr62_sp_energy} (a),
we see that the presence of the deformed $N$=38 gap is common to both cases.
The $N$=38 deformed gap at $\beta \sim$0.4 is universal 
in the sense that it exist both in the proton-rich and the neutron-rich regions.
This confirms that the deformation mechanism in the neutron-rich
 Cr and Fe isotopes is intimately related to that in the $N$=$Z\sim$38 region.

In spite of the above similarity, however, there is small but clear differences 
between the neutron Nilsson diagrams in the proton-rich $N$=$Z\sim$38 region and 
in the neutron-rich Cr-Fe region.
Looking closely at the $N$=38 deformed shell gap, we see that the 
relevant level crossing of the down-sloping $\nu g_{\frac{9}{2}}$ $\Omega=\frac{3}{2}$ orbit 
and the up-sloping orbit from $\nu f_{\frac{5}{2}}$ and $\nu p_{\frac{1}{2}}$ 
occurs at smaller deformation $\beta \sim$0.20 in the neutron-rich
 $^{62}$Cr than in $^{76}$Sr, where the crossing takes place at
 $\beta \sim$0.26.
It is also seen that the $N$=40 spherical gap at $\beta \sim$0 is smaller
 in the neutron-rich $^{62}$Cr than in $^{76}$Sr.
Both features are helpful to produce a largely deformed minimum at $\beta
\sim$0.27 in the deformation energy curve of $^{62}$Cr even though the deformation
driving effect due to the proton configuration is weaker in $^{62}$Cr than in $^{76}$Sr.
Note also that the deformed shell gap at $N$=40 around $\beta \sim 0.45$
present in $^{76}$Sr is barely seen in $^{62}$Cr. 
Another difference is also seen in the $N$=34 oblate gap which is large in
the proton-rich region (cf. Fig.\ref{Sr76_n_sp_energy}), leading to an oblate
ground state in $^{68}$Se ($\beta = -0.21$)\cite{yamagami,NAZ_NeZ,Fischer-Se}. 
However the $N$=34 oblate gap is much weaker in the neutron-rich Cr
isotopes(Fig.\ref{Cr62_sp_energy}(a)), 
and hence the deformation energy curve in $^{58}$Cr(Fig.\ref{Cr_isotope}) does
not show very strong tendency toward the oblate deformation. 

\subsection{Interaction dependence}
\begin{figure}[h]
  \begin{center}
     \includegraphics[width=8cm]{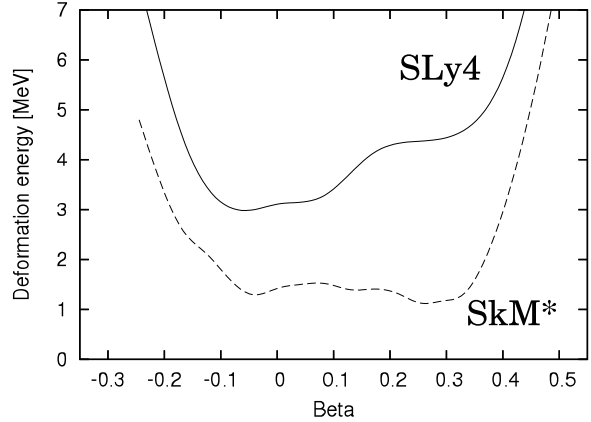} 
   \caption{The deformation energy curve obtained with SLy4 in
   $^{62}$Cr, plotted with the solid curve. The one for SkM* is also
   shown with the dotted curve for comparison.}\label{Cr62_SkMs_SLy4_energy}
  \end{center}
\end{figure}

As we have seen above, the Skyrme-HFB model using the parameter set SkM* 
produces the onset of large quadrupole deformation in the $N\sim$38-40
region of the Cr isotopes, being in qualitative agreement with the
experimental observations. 
It is noted however that the calculations using other parameter sets such
as SLy4, SkP\cite{THO_SkMs} and BSk2\cite{Goriely1} do not show largely
deformed minima in the same isotopes. 
Let us discuss how the onset of deformation depends on the Skyrme
parameter sets by comparing results which we obtain using SLy4 as well
as SkM*.
The calculation for SLy4 is the same as above except for the use of
different Skyrme parameters.
We show in Fig.\ref{Cr62_SkMs_SLy4_energy} the deformation energy 
curve in $^{62}$Cr obtained with the parameter set SLy4. 
The deformation energy curve for SLy4 does not have a well deformed
minimum, and it is stiff with respect to the quadrupole
deformation. 
We show the neutron Nilsson diagram for SLy4 in Fig.\ref{Cr62_sp_energy}
(c). 
By comparing with the Nilsson diagram for SkM* (Fig.\ref{Cr62_sp_energy}
(a)), 
we see a large difference in the position of the $\nu g_{\frac{9}{2}}$ 
orbit. The $N$=40 subshell gap at the spherical point is larger in the
case of SLy4 than SkM* by about 2 MeV. 
The higher position of $\nu g_{\frac{9}{2}}$ also shifts the position of
the $N$=38 deformed single-particle gap at 
larger deformation $\beta > 0.35$ in the SLy4 case.
These features apparently favor the stability of the spherical shape 
in contrast to the case of SkM*.
All these observations indicate that the position of the 
$g_{\frac{9}{2}}$ orbit plays a central role for the onset of the
quadrupole deformation in the neutron-rich Cr and Fe isotopes around $N$=38. 
The precise position of the $\nu g_{\frac{9}{2}}$ orbit and the size of the
$N$=40 subshell gap depend on the Skyrme parameter set. 
In other words, 
the deformation properties in the
neutron-rich Cr region provides us with a rather strong constraint for 
a proper choice of the Skyrme parameter set.

\section{Conclusions}
We have developed an axially symmetric Skyrme-HFB code based on the 2D mesh
representation in the cylindrical coordinate system in order to describe the 
quadrupole deformation of unstable nuclei.
Our code has enough accuracy to analyse the deformation property 
as tested by comparisons with the calculations using 
the THO and the B-Spline methods.

We applied the code to neutron-rich Cr, Fe and Ti isotopes around a
 possible new region of deformation with $N\gtrsim$38. 
The quadrupole deformation energy curve obtained with the Skyrme
 parameter set SkM* indicates an onset of deformation in the isotopes
 with $N\sim$38-42. 
The deformation energy curve is soft although the deformation of the
 minimum reaches $\beta \sim 0.25$ for $N$=38 and 40, 
and hence the nuclei in this region are expected to exhibit strong quadrupole
 collectivity having transitional character rather than that associated with
a well developed stable deformation. 
These results are in qualitative agreement with the trends of the presently available
 experimental data. 
The SkM* model describes also the Fe isotopes in the same mass region
 as transitional nuclei while the Ti isotopes are described as spherical
 nuclei having stiffer quadrupole deformation energy curve. 
By inspecting the neutron Nilsson diagram we discussed that the
 deformed $N$=38 gap emerging at $\beta \gtrsim 0.25$ as well as the position
 of $\nu g_{\frac{9}{2}}$ orbit play important role for the onset of
 quadrupole deformation. 
We have also shown that, contrary to SkM*, the parameter set SLy4 is not able
 to reproduce the onset of deformation. 
The sensitivity to the Skyrme parameter set arises from the fact that
 the deformation in this region emerges as a consequence of a delicate
 competition between spherical and deformed configurations, for which
 the position of $\nu g_{\frac{9}{2}}$ orbit plays an essential role.


\section*{Acknowledgements}
This work was supported by the Grant-in-Aid for Scientific
Research(No.17540244) from the Japan Society for the Promotion of
Science, and also by the JSPS Core-to-Core Program, International
Research Network for Exotic Femto Systems(EFES). 
The numerical calculations were carried out partly on SX8 at  YITP in Kyoto 
University.
%

\end{document}